\documentclass[12pt,twoside]{article}
\usepackage{amssymb}
\usepackage{amsmath,bbm}
\usepackage{blindtext}
\usepackage{titlesec}
\usepackage[makeroom]{cancel}
\usepackage{latexsym}
\usepackage{longtable}
\usepackage{epsfig}
\usepackage{graphicx,bbm,psfrag}
\graphicspath{{images/}}

\DeclareMathAlphabet{\mathsfit}{T1}{\sfdefault}{\mddefault}{\sldefault}
\SetMathAlphabet{\mathsfit}{bold}{T1}{\sfdefault}{\bfdefault}{\sldefault}
\usepackage{amsmath, amsfonts, amssymb, amsthm, amstext, bm, bbm, multirow, url}

\usepackage{lmodern}
\usepackage[T1]{fontenc}

\usepackage{graphicx, subcaption, ragged2e}
\DeclareCaptionJustification{justified}{\justifying}
\usepackage[ usenames, dvipsnames ]{xcolor}
\usepackage[colorlinks=true, linkcolor=blue,citecolor=red, urlcolor=blue, hypertexnames=true]{hyperref}

\usepackage[ textwidth=160mm, textheight = 210mm, hmarginratio={1:1}, vmarginratio={1:1}, 
			 heightrounded]{geometry} 

\begin{document}
\begin{titlepage}
\vspace*{30mm}
\begin{center}
\textbf{\Large{The Popkov-Sch\"{u}tz two-lane lattice gas:\medskip\\Universality for general jump rates}}\bigskip
\end{center}
\begin{center}
\large{Herbert Spohn}
\end{center}
\begin{center}
Technical University of Munich, 85747 Garching, Germany\\
Departments of Mathematics and Physics
\end{center}
 \vspace{30mm}
 \textbf{Abstract}. We consider the asymmetric version of the Popkov-Sch\"{u}tz two-lane lattice gas with general jump rates,
 subject to the stationary measure being of product form. This leaves still  five free parameters. At density $\tfrac{1}{2}$ the eigenvalues of the 
 flux Jacobian are degenerate. We compute the second order expansion of the fluxes  at density $\tfrac{1}{2}$
 and thereby identify the universality classes.\\\\
 December 30, 2025



\end{titlepage}
\newpage
\section{Introduction} 

There is renewed interest in nonreversible stochastic lattice gases  with two (or more) conserved components  in one dimension \cite{SKP25,RDMS25}. Due to nonreversibility, 
 a major obstacle is the availability of invariant measures. Firstly, in contrast with lattice gases satisfying detailed balance, there is no systematic method to determine such measures. In the case of nonreversible Glauber type dynamics, the generator of the Markov process has a spectral gap and perturbation theory can be used to study a wide range of invariant measures \cite{Liggett13}. However for lattice gases, because of conservation laws, 
a spectral gap is no longer available. Apparently, the only option is to rely on explicit solutions. 

For one component, usually available is the macroscopic current as a function of the density, $j(\rho)$. If  $j''(\rho) \neq 0$, one expects the
dynamic exponent $z = \tfrac{3}{2}$ and more generally KPZ scaling \cite{S16}. Of course, there are marginal cases with distinct behavior \cite{DLPS91}.  Numerically 
there is strong support confirming such predictions. There is also a variety of models for which KPZ scaling behavior is proved \cite{QS15,T18}.

For two components the understanding is much more partial. In our note we consider only models with exclusion rule, implying   $0 \leq\rho_1,  \rho_2 \leq 1$ for the two macroscopic densities. Expected is a well-defined current-density relation $j_m(\rho_1,\rho_2)$ yielding the flux Jacobian
matrix $A_{m,m'} = \partial j_m/\partial \rho_{m'}$, $m, m' = 1,2$. On general grounds the flux Jacobian has real eigenvalues. If the eigenvalues of $A$ are distinct, then the dynamic correlator has two sharp peaks moving with a velocity determined by the corresponding eigenvalue of $A$. These are the normal modes.
Since the peaks are separating linearly in time, one might guess that their behavior is similar to the one-component case.  While the complete picture is 
somewhat more involved \cite{S14,PS15}, the available numerical support is convincing.

In our article we will discuss the case when the flux Jacobian has a degenerate spectrum. Then the two modes are sitting on top of each other 
thereby  interacting strongly. To understand the resulting dynamical behavior is more difficult. Actually, there is a more mundane obstacle. Generically, the two eigenvalues of $A$ are distinct for the full range of densities. It requires fine-tuning of the model parameters, including both densities, to achieve degeneracy.
Some time back Popkov and Sch\"{u}tz  \cite{PS03}  proposed an accessible model, which consists of two lanes of particles. More pictorially the position space is a ladder. Particles jump only in their own lane respecting the exclusion rule. For decoupled lanes, the top lane is ASEP and so is the bottom lane. Coupling is introduced 
through  jump rates depending on the configuration in the opposite lane, while particles always stay in their own lane. 
To enable degeneracy, the rates for the bottom  lane are assumed to be the space reflected jump  rates of the top lane.  In \cite{PS03} this is the so-called asymmetric version. Macroscopically reflection symmetry leads to the constraint  
\begin{equation}
    \label{1}
 j_1(\rho_1,\rho_2) = - j_2(\rho_2, \rho_1).
 \end{equation}
 
 In \cite{PS03} a condition is discovered, under which the steady state is of product form with respect to rungs. 
The TASEP version of this model was then investigated in \cite{PS12} with the result that exactly at densities $\rho_1 = \tfrac{1}{2} = \rho_2$ the flux Jacobian  $A$ has a degenerate eigenvalue $0$. Away from these densities, $A$ is nondegenerate. To achieve such property the jump rates are fixed up to a single free  parameter, denoted by $\gamma >0$. Much later this particular case was studied numerically  in \cite{SKP25,RDKKS24} arriving at two main results: (i)  The dynamical exponent seems to depend on $\gamma$.
Distinct values of $\gamma$ define distinct universality classes, since the scaling function depends nontrivially, i.e. beyond  dilations and rotations, on $\gamma$.

The model of two coupled  ASEP satisfying the reflection symmetry is studied in this article. To  have  stationary product measures leaves five free parameters. $\rho_1 = \tfrac{1}{2} = \rho_2$ is still the only point of degeneracy. We follow the strategy in \cite{RDKKS24}: 
The steady state current is expanded to second order around $\rho_1= \tfrac{1}{2}= \rho_2$. As it should be, the linear term vanishes. The quadratic term is rotated by $\pi/4$, which then determines  the universality class, see the discussion in \cite{RDKKS24}.
\section{Popkov-Sch\"{u}tz lattice gas}
The occupation numbers of the top lane, also lane 1, are denoted by $\{\eta_j, j \in \mathbb{Z}\}$ and the ones   of the bottom lane, also lane 2,  by $\zeta_j$. For lane  $1$ one has  the usual nearest neighor exchanges between $\eta_j$ and $\eta_{j+1}$ with a rate which  depends on $\zeta_j,\zeta_{j+1}$, as formula
\begin{eqnarray}
    \label{2}
 &&\hspace{-25pt}c_{j,j+1}^\eta = \eta _j(1-\eta_{j+1})\big( \alpha (1-\zeta_j)(1 -  \zeta_{j+1}) + \beta  \zeta_j \zeta_{j+1} + \gamma  \zeta_j(1 -  \zeta_{j+1})
 + \delta (1-\zeta_j)\zeta_{j+1}\big)\nonumber\\[1ex] 
 &&\hspace{-15pt} + (1 -\eta_j)\eta_{j+1}\big(\tilde{\alpha} (1-\zeta_j)(1 -  \zeta_{j+1}) + \tilde{\beta}  \zeta_j \zeta_{j+1} + \tilde{\gamma}  \zeta_j(1 -  \zeta_{j+1}) 
 + \tilde{\delta} (1-\zeta_j)(\zeta_{j+1})\big),
\end{eqnarray}
where the parameters satisfy $\alpha, \beta,\gamma,\delta \geq 0 $ and $\tilde{\alpha}, \tilde{\beta},\tilde{\gamma}, \tilde{\delta} \geq 0 $. The exchange rates for lane 2 are determined by spatial reflection and are thus given by
\begin{eqnarray}
    \label{3}
 &&\hspace{-25pt}c_{j,j+1}^\zeta = \zeta_j(1-\zeta_{j+1})\big( \tilde{\alpha} (1-\eta_j)(1 -  \eta_{j+1}) +  \tilde{\beta}  \eta_j \eta_{j+1} +  \tilde{\delta}  \eta_j(1 -  \eta_{j+1}) 
 +  \tilde{\gamma} (1-\eta_j)\eta_{j+1}\big)\nonumber\\[1ex]
 &&\hspace{-15pt} + (1- \zeta_j)\zeta_{j+1}\big(\alpha (1-\eta_j)(1 -  \eta_{j+1}) +\beta  \eta_j \eta_{j+1} + \delta  \eta_j(1 -  \eta_{j+1}) 
 + \gamma (1-\eta_j)\eta_{j+1}\big).
\end{eqnarray}

We want the steady state to be product with respect to $j$. Thus the single rung probability is taken as
\begin{equation}
    \label{4}
 Z^{-1}\exp\big(-\nu\eta_j\zeta_j + \mu_1 \eta_j +\mu_2 \zeta_j\big).
    \end{equation}
Indeed, as established in \cite{PS03}, this product measure is stationary provided that
\begin{equation}
    \label{5}
    \alpha - \tilde{\alpha} = \beta - \tilde{\beta} = \gamma \mathrm{e}^{-\nu} - \tilde{\gamma} = \delta - \tilde{\delta}  \mathrm{e}^{-\nu}. 
    \end{equation}
Here $\mu_1,\mu_2$ are the chemical potentials, controlling the densities, and $\nu$ is a stationary state parameter controlling the correlations between lanes. Setting $\rho_1= \langle \eta_j \rangle =u$ and  $\rho_2 = \langle \zeta_j \rangle =v$, average according to \eqref{4},  the four elementary probabilities of a single rung are  expressed as
\begin{equation}
    \label{6}
    p_{00} = 1 -u-v +\Omega, \quad p_{01} = v - \Omega,\quad p_{10} = u- \Omega, \quad p_{11}  = \Omega, 
\end{equation}
where
\begin{equation}
    \label{7}
    \Omega =\frac{1}{2q}\Big( -1- q(1-u-v) + \sqrt{(1+q(1-u-v))^2 +4quv }\Big)
\end{equation}
and 
\begin{equation}
    \label{8}
  q =   \mathrm{e}^{\nu} -1. 
   \end{equation}

The steady state current of lane $1$ is given by
\begin{eqnarray}
    \label{9}
 &&\hspace{-30pt} j_1(u,v)  = \langle  c_{j,j+1}^\eta \rangle \\[1ex]
 &&\hspace{-15pt}= a(1 -u-v+\Omega)(u-\Omega) + a (v-\Omega)\Omega + c (1-u-v + \Omega) + d (u- \Omega)(v-\Omega),\nonumber
   \end{eqnarray}
 using the shorthand
\begin{equation}
    \label{10}
 a = \alpha - \tilde{\alpha}, \quad    b = \beta - \tilde{\beta}, \quad    c = \gamma - \tilde{\delta}, \quad    d = \delta - \tilde{\gamma}.
    \end{equation}
Note that  $a = b$ by stationarity. For the average current of the second lane,  there is a similar expression. But it suffices to note that by reflection symmetry
\begin{equation}
    \label{11}
    j_1(u,v) = -  j_2(v,u).\medskip
    \end{equation}
 \section{Expansion of the steady state current} The goal is to carry out a second order expansion of the current \eqref{9} at $u= \tfrac{1}{2}$, $v=\tfrac{1}{2}$, in analogy to \cite{RDKKS24}.     
  Since the steady state is uncorrelated in the lattice index, one expects to end up with cyclic parameters for the quadratic terms. It is still of interest, to explore how the required algebra works, in particularly to thereby determine the respective universality class.
  
  The key step is to note the identity 
  \begin{equation}
    \label{12}
   j_1(u,v) = au(1-u) + (d-a)(u-\Omega)(v-\Omega) +  (c-a) (1-u-v + \Omega)\Omega.
    \end{equation}
 We expand as  
  \begin{equation}
    \label{13}
    u = \tfrac{1}{2} + \epsilon, \quad  v = \tfrac{1}{2} + \kappa. 
    \end{equation}  
   Then  
    \begin{eqnarray}
    \label{14}
 &&\hspace{-25pt}(u-\Omega)(v-\Omega) = (\tfrac{1}{2} +\epsilon -\Omega)(\tfrac{1}{2} +\kappa - \Omega) \\
 &&\hspace{-15pt}= \frac{1}{4q^2} \big(1 + q - \sqrt{1+q}\big)^2 - \tfrac{1}{4} \sqrt{1+q}(\epsilon^2 +\kappa^2) + \epsilon \kappa \big( 1 + q^{-1} + \tfrac{1}{2} \sqrt{1+q} - q^{-1}\sqrt{1+q}\big).\nonumber
    \end{eqnarray}
    In addition,
      \begin{eqnarray}
    \label{15}
 && \hspace{-40pt}(1-u-v + \Omega)\Omega = (1 - \tfrac{1}{2} -\epsilon - \tfrac{1}{2} -\kappa +\Omega)\Omega \\
 &&\hspace{-15pt}= \frac{1}{4q^2} \big( \sqrt{1+q} -1\big)^2 - \frac{1}{4 \sqrt{1+q}} (\epsilon^2 +\kappa^2) + \epsilon \kappa 
 \big( \frac{1}{q}- \frac{1}{q\sqrt{1+q}} - \frac{1}{2\sqrt{1+q}} \big).\nonumber
    \end{eqnarray}
The first order currents vanish, thereby confirming that  the flux Jacobian vanishes at $(\tfrac{1}{2},\tfrac{1}{2})$.  The second order reads
  \begin{eqnarray}
    \label{16}
 &&\hspace{-20pt} j_{1,2} (\epsilon, \kappa) = - a \epsilon^2 
  + (d-a)\big(- \tfrac{1}{4}\sqrt{1+q}(\epsilon^2 +\kappa^2) + \epsilon\kappa\big(1 + q^{-1} - \tfrac{1}{2} \sqrt{1+q}-q^{-1}\sqrt{1+q}\big)\big)\nonumber\\ 
 &&\hspace{40pt} + 
 (c-a)\big( - \frac{1}{4 \sqrt{1+q}} (\epsilon^2 +\kappa^2) + \epsilon \kappa 
 \big( \frac{1}{q}- \frac{1}{q\sqrt{1+q}} - \frac{1}{2\sqrt{1+q}} \big).
 \end{eqnarray} 
 
 Without loss of generality, we can choose the time scale such that $a=1$. Let us denote the coupling constants by $x = c-1$ and $y= d -1$. Then 
   \begin{equation}
    \label{17}
  x = \gamma - \tilde{\delta} -1,\quad y = \delta - \tilde{\gamma} -1.
    \end{equation} 
 Stationarity implies
   \begin{equation}
    \label{18}
   \gamma \mathrm{e}^{-\nu} - \tilde{\gamma} = 1,\quad  \delta - \tilde{\delta}\mathrm{e}^{-\nu} = 1.
    \end{equation}    
Using $1+q =   \mathrm{e}^{\nu} $, one concludes that  
  \begin{equation}
    \label{19}
   x = q - (1+q)y.    
   \end{equation}     
  Upon inserting   in \eqref{16}, the two terms proportional to $y$  cancel each other.  Hence, restoring general $a$, the second order stationary current simplifies to
    \begin{equation}
    \label{20}
 j_{1,2} (\epsilon, \kappa) = a\big(-  \epsilon^2 
   - \frac{q}{4 \sqrt{1+q}} (\epsilon^2 +\kappa^2) +
 \big( \frac{2 \sqrt{1+q} -2 -q}{2\sqrt{1+q}}\big) \epsilon \kappa   \big).
 \end{equation} 

The parameters $b,c,d$ have dropped out and $a$ is a mere time scale. Therefore \eqref{20} has to be  valid also for TASEP with the set of parameters studied in \cite{RDKKS24}. In this case jumps are only one-sided, i.e. the tilde parameters vanish, 
$a= \alpha = \beta = \delta = 1$, and $\gamma = 1+q$. Indeed \eqref{20} passes the test. 
\section{Summary and discussions}
 Our result is surprisingly simple. The parameter $q > -1$ controls the steady state through $q =  \mathrm{e}^{\nu} -1$. As first discussed in \cite{RDKKS24}, and confirmed in \cite{SKP25} and \cite{RDMS25}, the parameter  $q$ also labels the universality classes. More precisely,   the two density fields are rotated by $\pi/4$, which makes the transformed fields to be uncorrelated. But one still  has to normalize their variance to $1$. Then the continuum approximation to second order is given by  the two coupled Burgers equations
 \begin{equation}\label{21}
    \begin{aligned}
	\partial_t \phi_1 &= \partial_x \big(  2  \phi_1 \phi_2 + \tfrac{1}{2} \partial_x \phi_1+ \xi_1 \big), \\
	\partial_t \phi_2 &= \partial_x \big(  \phi_1^2 + (2-(1+q)^{-\frac{1}{2}})\phi_2^2 +  \tfrac{1}{2}\partial_x \phi_2  +\xi_2 \big).
    \end{aligned}
\end{equation}  
Here $\xi_1,\xi_2$ are two independent unit spacetime white noise processes. As worked out in detail in \cite{RDMS25}, after a suitable  rescaling, these equations can be written also as 
 \begin{equation}\label{22}
    \begin{aligned}
	\partial_t \phi_1 &= \partial_x \big(  2 X \phi_1 \phi_2 + \tfrac{1}{2} \partial_x \phi_1+ \xi_1 \big), \\
	\partial_t \phi_2 &= \partial_x \big(  X\phi_1^2 + \phi_2^2 +  \tfrac{1}{2}\partial_x \phi_2  +\xi_2 \big)
    \end{aligned}
\end{equation}  
with
\begin{equation}
\label{23}
X = \frac{\sqrt{1+q}}{2\sqrt{1+q} -1}.
\end{equation} 

For the coupled Burgers equations studied in \cite{RDKKS24}, the first equation is the same as \eqref{22}, but in the second equation the nonlinearity is more general and reads  $Y\phi_1^2 + \phi_2^2$. The studied parameter space is the $X$-$Y$ plane with $XY >0$. The diagonal $\{X=Y\}$ plays a special role. The corresponding coupling matrices are cyclic. Along the diagonal the scaling functions of the dynamic correlator change nontrivially \cite{RDKKS24} and presumably the dynamical exponent varies continuously with $X$, still being  close to $\tfrac{3}{2}$ \cite{SKP25}.
In this parametrization the studied lattice gas is located on the diagonal $\{X=Y\}$. As $q$ changes,   the parameter $X$ varies along the diagonal. However the interval $0 \leq X \leq \tfrac{1}{2}$ is not covered.

Now given $q$, there are eight free parameters available, $\alpha, \beta,\gamma, \delta$ and $ \tilde{\alpha},  \tilde{\beta},  \tilde{\gamma}, \tilde{\delta}$. They have to satisfy the three identities \eqref{5}. Furthermore, one can choose a global time scale. 
This yields a universality class with four independent parameters. For TASEP, the tilde parameters vanish and the class consists of a single point. 
 
 For TASEP the only cyclic choice is $\alpha = \beta = \delta = 1$ and arbitrary $\gamma >0$. Moving to other parameters the lattice gas becomes noncyclic. The steady state has to be studied numerically. For sure independence is lost. But based on previous experience the correlations in the steady state should still have a rapid decay. It would be worthwhile to study such larger class.
 
 In case of TASEP, interchanging particles and holes for both lanes is equivalent to interchanging the two components. In the dynamical correlator, $S(j,t)$, this particle-hole symmetry is reflected by the identities $S_{11} = S_{22}$ and $S_{12} = S_{21}$. For ASEP such  symmetry would require $\alpha = \beta$
and $\tilde{\alpha} =  \tilde{\beta}$. On the other side, in the continuum approximation,  particle-hole symmetry is replaced by the symmetry $\phi \to -\phi_1$. Rotating back this information to the ASEP lattice gas, one would expect that, while violated, the identities   $S_{11} = S_{22}$, $S_{12} = S_{21}$ are regained in the scaling limit.\\\\
\textbf{Acknowledgments}. I thank Abhishek Dhar and Manas Kulkarni for most instructive discussions and the hospitality at the ICTS. For support, I am grateful to  the VAJRA faculty scheme (No. VJR/2019/000079) from the Science and Engineering Research Board (SERB), Department of Science and Technology, Government of India. 
     
\end{document}